\documentclass[preprintnumbers,amsmath,11pt,amssymb,floatfix,superscriptaddress,nofootinbib]{article}

\def\mytitle#1{\setcounter{equation}{0}
\setcounter{footnote}{0}
\begin{flushleft}\Large\textbf{#1}\end{flushleft}
\vspace{0.25cm}}
\def\myname#1{\leftline{{\large #1}}\vspace{-0.13cm}}
\def\myplace#1#2{\small\begin{flushleft}\textit{#1}\\
\texttt{#2}\end{flushleft}}

\def\myclassification#1{\small\noindent
PACS:
       #1\vspace{0.5cm}}
\usepackage{graphicx}% Include figure files
\begin{document}

\mytitle{Dark Energy Accretion onto Van der Waal's Black Hole}

\myname{ $Sandip~Dutta$\footnote{duttasandip.mathematics@gmail.com} and $~Ritabrata~Biswas$\footnote{biswas.ritabrata@gmail.com}}
\myplace{Department of Mathematics, The University of Burdwan, Golapbag Academic Complex, Burdwan-713104, Dist: Purba Barddhaman, Sate: West Bengal, India.}{} 
 
%%%%%%%%%%%%%%%%%%%%%%%%%%%%%%%%%%%%%%%%%%%%%%%%%%%%%%%%%%%%%%%%%%%%%%%%%%%%%%%%%%%%%%%%%%%%%%%%%%%%%%%%%%%%%%%%%%%%%%
\begin{abstract}
We consider the most general static spherically symmetric black hole metric. The accretion of the fluid flow around the Van der Waal's black hole is investigated and we calculate the fluid's four-velocity, the critical point and the speed of sound during the accretion process. We also analyze the nature of the universe's density and the mass of the black hole during accretion of the fluid flow. The density of the fluid flow is also taken into account. We observe that the mass is related to redshift. We compare the accreting power of the Van der Waal's black hole with Schwarzschild black hole for different accreting fluid.
 \end{abstract}
%%%%%%%%%%%%%%%%%%%%%%%%%%%%%%%%%%%%%%%%%%%%%%%%%%%%%%%%%%%%%%%%%%%%%%%%%%%%%%%%%%%%%%%%%%%%%%%%%%%%%%%%%%%%%%%
%{\bf Keywords:} Black hole physics, Thermodynamics,  Accretion disc, Extended Chaplygin gas.\\
\myclassification{ 04.70.Dy, 95.30.Sf, 95.35.+d, 95.36.+x, 98.80.Cq, 98.80.-k }

%\newpage
%%%%%%%%%%%%%%%%%%%%%%%%%%%%%%%%%%%%%%%%%%%%%%%%%%%%%%%%%%%%%%%%%%%%%%
\section{Introduction}
On the Anti de-Sitter(AdS) boundary, the studies of strongly coupled thermal field theories led us to some interesting results regarding the physics of asymptotically AdS black holes. In the reference \cite{hawking}, for Schwarzschild-AdS black hole space-time, a first order phase transition of thermal radiation/black hole is observed. The thermodynamic behaviours make after those of Van der Waal's fluid if charge and rotation are incorporated \cite{2,3,4,5}. A more prominent analogy \cite{6,7} is observed when cosmological constant is treated as a thermodynamic pressure, $P$, given as
\begin{equation}
P=-\frac{\Lambda}{8 \pi}=\frac{3}{8 \pi l^2}~~.
\end{equation}
This also extends the first law of black hole thermodynamics as $\delta M =T \delta S +V \delta P+ ...$. To do so we need to introduce a quantity which is thermodynamically conjugated to $P$ and  is interpreted as a black hole thermodynamic volume \cite{8,9} which is given as
\begin{center}
$V=\left( \frac{\partial M}{\partial P}\right)_S$~~~~.\\
\end{center}
Immediately after the consideration of analogical pressure and volume, one may query about the nature of black hole's equation of state. Comparison between the temparetures of the black holes and concerned fluid or the comparison between the volumes or the pressures of the black holes to the corresponding physical properties of concerned fluid (with which the black holes' equation of state does resemble) can be built up. References like \cite{Vanderwaal} have chosen the equation of state possessed by Van der Waal's fluids. For one mole of such fluid, the equation of state turns to be
\begin{equation}
T=\left( P + \frac{a}{v^2}\right) \left( v-b \right),
\end{equation}
where $v=\frac{V}{N}$, $N$ being the fluid's degrees of freedom. The attraction between two molecules of the concerned fluid is measured by the parameter $'a'$($a>0$). On the other hand, the volumetric measure is kept by the parameter $'b'$. It is studied in the references \cite{6,7} that the rotating (or nonrotating as well) AdS black holes' thermodynamic natures resemble with this particular type of fluid to a large extent. Whenever a phase transition is observed in the black hole, a swallowtail catastrophe is  exposed by Gibb's free energy for both the fluid and the black hole. An exact match between the properties of Van der Waal's fluid and a particular type of black holes are tried to be obtained in the references \cite{Debnath11,Debnath12}. The static spherically symmetric black hole metric obtained by these article is given as
\begin{equation}\label{s1}
ds^2= -f(r)dt^2 + \frac{dr^2}{f(r)}+ r^2(d\theta^2 + \sin^2 \theta d\phi^2)'
\end{equation} 
with the lapse function 
\begin{equation}\label{s2}
f(r)= 2\pi a_{vw}- \frac{2M}{r} + \frac{r^2}{l^2}\left(1+\frac{3b_{vw}}{2r}\right)-\frac{3\pi a_{vw} b_{vw}^2}{r(2r+3b_{vw})} - \frac{4\pi a_{vw} b_{vw}}{r}\log{\left(\frac{r}{b_{vw}}+\frac{3}{2}\right)},
\end{equation}
where $M$ is the mass of the Van der Waal's black hole. For a small scale connected to the inverse cosmological constant $\Lambda$, $l$ is a parameter with dimensions of length (Hubble length) and the parameter $a_{vw}>0$ measures the attraction between the molecules of the fluid, and the parameter $b_{vw}$ measures their volume.
We find several literatures, nowadays, where \cite{Cadoni,Saurav} the thermodynamic natures of the Van der Waal's black holes have been studied. A common conclusion drawn from all these articles is that the Van der Waal's black hole solution is qualitatively analogical (on a thermodynamic perspective) with the Van der Waal's fluid. Else we can treat it as just another type of static spherically symmetric black hole ansatz.

As we know, since almost twenty years from now, that our universe is experiencing a late time cosmic acceleration and to justify such kind of accelerated expansion, we have proposed many models, where a homoganeous exotic fluid coined as `quintessence', `dark energy' or `phantom field' etc is assummed to be present in the universe (spreaded all over in it) and is exerting negative pressure.

Until now, many dark energy models have been speculated. Between these models, the most appealing one is the cosmological constant \cite{Padmanabhan1} which is the simplest candidate of dark energy, distinguished by the equation of state $p=\omega \rho$ with $\omega=-1$. However, two problems raise for the cosmological constant model : the coincidence problem \cite{Copeland1} and the fine tuning problem. Some methods are given to solve these problems, such as considering the holographic principle \cite{Hooft1}, anthropic principle \cite{Wilczek1}, invoking an interaction between dark matter and dark energy\cite{Del1} and variable cosmological constant scenario\cite{Jamil1}. In this context, several well-known models such as phantom, quintom, Chaplygin gas, quintessence, geographic dark energy and holographic dark energy, etc., have been proposed\cite{Wetterich1, Caldwell1, Elizalde1, Wu1, Li1, Cai1, Li2, Li3}. The Chaplygin gas is an exotic type of fluid whose energy density $\rho$ and pressure $p$ fit the equation of state $p=-B/\rho$, where $B$ is a positive constant \cite{Kamenshchik1}. At large value of scale factor, the Chaplygin gas tends to accelerate the universe's expansion, however, at small value of the scale factor it acts as pressureless fluid. 

In general terms, when such exotic fluids violate the strong energy condition $3p+\rho >0$, we call them to be in the quintessence era. But as they violate the weak energy condition, i.e., $p+\rho >0$ we say that these fluids are in phantom era. In phantom era, a future singularity named ``Big Rip" may occur where all the four fundamental forces might be defeated by the dark energy and every matter will be destroyed. Now it was a challenging question that the compact objects like black holes will exist at big rip or not. Babichev et al \cite{Babichev2004 DE accretion PRL} for the first time in literature, have general relativistically studied the phantom energy accretion on Schwarzschild black hole and shown that if the black hole is surrounded by a fluid which is following the equation $p+\rho<0$ then the mass of the black hole will be decreased by the effect of such kind of fluid's accretion. Dark energy accretion and its properties for different black holes as central engine and different equation of states of accreting fluid are studied in different references\cite{Jamil1,Jamil2,Jamil3,Jamil4,Biswas1,Huang1,Debnath1}.

The general overview is that the black holes will loss mass due to such kind of exotic matter accretion. A series of pseudo Newtonian studies of accretion disc nature for dark energy accretion are done in references \cite{Biswasaccretion1,Biswasaccretion2,Sandip1,Sandip2}. It is speculated that dark energy accretion strengthens the wind branch and weakens the accretion branch which as a result faints the accretion disc, i.e., it weakens the feeding process of the black holes. Effects of different modified gravity parameters on accretion are interesting topics to study.

Van der Waal's black hole is a thermodynamically modified version of static spherically symmetric black hole which incorporates extra parameters $a_{vw}$ and $b_{vw}$ which modify the black hole's gravitating power. As a fluid, Van der Waal's gas incorporats the mutual attractions between molecules and the pressure exerted by molecules on the container's wall. It is expected that the Van der Waal's black hole will take into account  the mutual interactions of microstates in it. This mutual attraction might have effect on the external power of the black hole. Our motivation for this paper is to study how does the black hole accretion disc behave if both the central black hole is modified as a Van der Waal's black hole and the accreting fluid is taken to be an exotic one. This may give rise to a new construction of black hole's mass reduction formula. Besides the nature of the mass of the black holes, the density variation near the black hole will be studied. This may speculate how much of negative pressure generating nature is carried out by the dark energy models near this particular type of compact objects. 

In the present paper, we consider most general static spherically symmetric black hole solution in section $2$. An investigation regarding the accretion of any general kind of fluid flow around the black hole is done in the same section. Next, we analyze the accretion of the fluid flow around Van der Waal's black hole in section $3$. Here we calculate the existence(s) of critical point(s), velocity of sound and the fluid's four velocity during the process of accretion. Finally, we briefly  conclude through a discussion.

\section{Accretion onto a General Static Spherically Symmetric Black Hole}

We will consider a general static spherically symmetric metric \cite{Debnath11, Debnath12} written as equation (\ref{s1}), where $f(r)(>0)$ considered as a function of $r$  and $M$ as the mass of the black hole.

For the accreting fluid, energy-momentum tensor is given by ($8\pi G=c=1$)
\begin{equation}\label{2}
T_{\mu\nu}=(\rho+p)u_\mu u_\nu+pg_{\mu\nu}~~~,
\end{equation}  
where $p$ and $\rho$ are the pressure and energy density of the fluid. Also, $u^\mu = \frac{dx^\mu}{ds} = (u^0, u^1, 0, 0)$ is the four-velocity vector of the fluid flow, where $u^0$ and $u^1$ are the components (non-zero) of velocity vector satisfying 

$u_\mu u_\nu=-1$   $\Rightarrow g_{00} u^0 u^0 + g_{11} u^1 u^1=-1$   $\Rightarrow (u^0)^2 =\frac{f+(u^1)^2}{f^2}$.

Let us consider the radial velocity of the flow $u^1=u$. Therefore, $u_0=g_{00} u^0 = \sqrt{u^2 +f}$ where $\sqrt{-g}=r^2 sin\theta$.

From the equation (\ref{2}), we get $T^1_0 =(\rho+p)u_0 u$. For inward flow, assuming $u < 0$ (as the fluid flows towards the black hole). 

For the fluid flow, we may take that the fluid is any kind of dark energy or dark matter. When a static spherically symmetric black hole is considered, a proper dark-energy accretion model should be gained by generalizing Michel's theory\cite{Michel1}. Babichev et al.\cite{Babichev1, Babichev2} have performed the generalization of the dark energy accretion onto Schwarzschild black hole. $T^{\mu\nu}_{;\nu}=0$ is the energy-momentum conservation law for the relativistic Bernoulli equation (the time component). 

Taking the radial temporal component of relativistic energy momentum conservation equation, we have  $\frac{d}{dr}(T^1_0\sqrt{-g})=0$, which provides the first integral $(\rho+p)u_0 u^1\sqrt{-g}={\cal C}_1$. Hence, 
\begin{equation}\label{3}
ur^2(\rho+p)\sqrt{u^2+f}={\cal C}_1 ~~~~~,
\end{equation} 
where ${\cal C}_1$ is an integrating constant, having the dimension of the energy density. For the energy momentum tensor, the energy flux equation can be defined by the projection of the conservation law, i.e., $u_\mu T^{\mu\nu}_{;\nu} =0 \Rightarrow u^{\mu} \rho_{,\mu}+(\rho+p)u^{\mu}_{,\mu} = 0$. From this, we get (taking $\mu = 1$), 
\begin{equation}\label{4}
ur^2exp\left[\int_{\rho_\infty}^{\rho_h}\frac{d\rho}{\rho+p(\rho)}\right]= -{\cal C}~~~~~,
\end{equation}  
where ${\cal C}$ is a constant of integration constant and for convenience the minus sign is taken. Moreover, $\rho_\infty$ and $\rho_h$ denote the the energy densities at infinite distance from the black hole and at the black hole horizon respectively. From equation (\ref{3}) and (\ref{4}), we get, 
\begin{equation}\label{5}
(\rho+p) exp\left[-\int_{\rho_{\infty}}^{\rho_h}\frac{d\rho}{\rho+p(\rho)}\right]\sqrt{u^2+f}= {\cal C}_2~~~~~,
\end{equation}
where ${\cal C}_2 =-{\cal C}_1/{\cal C} = \rho_\infty+ p(\rho_\infty)$. Also $J^{\mu}_{;\mu} = 0$ is the equation of mass which gives $\frac{d}{dr}(J^1\sqrt{-g})= 0 \Rightarrow \rho u^1 \sqrt{-g} = A_1$ and yields
\begin{equation}\label{6}
\rho u r^2={\cal C}_3~~~~~~,
\end{equation}
where ${\cal C}_3$ is another integrating constant. From (\ref{3}) and (\ref{6}), we get,
\begin{equation}\label{7}
\frac{\rho+p}{\rho}\sqrt{u^2+f} = \frac{{\cal C}_1}{{\cal C}_3} = {\cal C}_4 = constant.
\end{equation}  
Let us assume
\begin{equation}\label{8}
V^2 = \frac{d\ln{(\rho+p)}}{d \ln{\rho}} - 1
\end{equation}
From the equation (\ref{6}), (\ref{7}) and (\ref{8}), we get,
\begin{equation}\label{9}
\left[V^2-\frac{u^2}{u^2+f}\right]\frac{du}{u} + \left[-2V^2 + \frac{rf'}{2(u^2+f)}\right]\frac{dr}{r} = 0 
\end{equation}
If one or the other of the bracketed factors in (\ref{9}) is terminated, we obtain a turn-around point and for this case, the solutions will give two values in either $r$ or $u$. The solutions are passing through a critical point that assembles the material falling out (or flowing into) and along with the particle trajectory the object has monotonically increasing velocity. Critical point is a point where the speed of the flow is equal to the speed of sound inside the fluid. Assuming at $r = r_c$, where the critical point of accretion is located, which can be obtained by assuming the two bracketed terms (the coefficients of $dr$ and $du$) in equation (\ref{9}) to be zero. Therefore, at $r = r_c$, we get,
\begin{equation}\label{10}
V_c^2 = \frac{u_c^2}{u_c^2+f(r_c)}~~and~~\frac{4V_c^2}{r_c} = \frac{f'(r_c)}{u^2+f(r_c)}~~,
\end{equation}
where $u_c$ is the critical speed of the flow at $r = r_c$ (at the critical point) and the subscript $c$ is denoting the critical value. From (\ref{10}), we get,
\begin{equation}\label{12}
u_c^2 = \frac{r_cf'(r_c)}{4}
\end{equation}
and 
\begin{equation}\label{13}
V^2_c = \frac{r_cf'(r_c)}{4f(r_c)+r_cf'(r_c)}~~.
\end{equation}
At $r=r_c$, the sound speed can be obtained by
\begin{equation}\label{14}
c_s^2 = \frac{dp}{d\rho} \Big\arrowvert_{r=r_c}   = \frac{{\cal C}_4V_c(V_c^2+1)}{u_c}-1
\end{equation} 
The solutions are physically admissible if $u_c^2>0$ and $V_c^2>0$.
\section{Accretion onto a Van der Waal's Black Hole}
%%%%%%%%%%%%%%%%%%%%%%%%%%%%%%%%%%%%%%%%%%%%%%%%%%%%%%%%%%%%%%%%%%%%%%

Considering that the fluid flow accretes upon the Van der Waal's black hole, we will compute the expressions of $u_c^2,~ V_c^2$ and $c_s^2$ at $r=r_c$ (i.e., at the critical point). We get (using equations (\ref{12}) and (\ref{13})):
$$u_c^2 = \frac{r_c}{4} \Bigg[ -\frac{3b_{vw}}{2l^2}+\frac{2M}{r^2_c}+\frac{2\left(1+\frac{3b_{vw}}{2r_c}\right)r_c}{l^2}+ \frac{6 a_{vw} b_{vw}^2 \pi}{r_c(2r_c+3b_{vw})^2}+\frac{3\pi a_{vw} b_{vw}^2}{r_c^2(2r_c+3b_{vw})}-\frac{4a_{vw} \pi}{r_c\left(\frac{r_c}{b_{vw}}+\frac{3}{2}\right)}~~~~~~~~~~~~~~~~~~~~~~~~~~~~~~~~~~~$$
\begin{equation}\label{16}
~~~~~~~~~~~~~~~~~~~~~~~~~~~~~~~~~~~~~~~~~~~~~~~~~~~~~~~~~~~~~~~~~~~~~~~~~~~~~~~~~~~~~~~~~~~~~~~~~~~~+\frac{4 a_{vw} b_{vw} \pi \log{\left(\frac{3}{2}+\frac{r_c}{b_{vw}}\right)}}{r^2_c}\Bigg],
\end{equation}
\begin{equation}\label{17}
V^2_c =  \Bigg\{ 1+\frac{4}{r_c}\frac{ 2\pi a_{vw}- \frac{2M}{r_c} + \frac{r_c^2}{l^2}\left(1+\frac{3b_{vw}}{2r_c}\right)-\frac{3\pi a_{vw} b_{vw}^2}{r_c(2r_c+3b_{vw})} - \frac{4\pi a_{vw} b_{vw}}{r_c}\log{\left(\frac{r_c}{b_{vw}}+\frac{3}{2}\right)}}{-\frac{3b_{vw}}{2l^2}+\frac{2M}{r^2_c}+\frac{2\left(1+\frac{3b_{vw}}{2r_c}\right)r_c}{l^2}+ \frac{6 a_{vw} b_{vw}^2 \pi}{r_c(2r_c+3b_{vw})^2}+\frac{3\pi a_{vw} b_{vw}^2}{r_c^2(2r_c+3b_{vw})}-\frac{4a_{vw} \pi}{r_c\left(\frac{r_c}{b_{vw}}+\frac{3}{2}\right)}+\frac{4 a_{vw} b_{vw} \pi \log{\left(\frac{3}{2}+\frac{r_c}{b_{vw}}\right)}}{r^2_c}} \Bigg\}^{-1}
\end{equation}
and $c_s^2$ can be obtained by using the equations (\ref{14}), (\ref{16}) and (\ref{17}).

The physically admissible solutions of the above equations are obtained if $u^2_c > 0$ and $V_c^2 > 0$,  i.e.,
$$\frac{2M}{r^2_c}+\frac{2\left(1+\frac{3b_{vw}}{2r_c}\right)r_c}{l^2}+ \frac{6 a_{vw} b_{vw}^2 \pi}{r_c(2r_c+3b_{vw})^2}+\frac{3\pi a_{vw} b_{vw}^2}{r_c^2(2r_c+3b_{vw})}~~~~~~~~~~~~~~~~~~~~~~~~~~~~~~~~~~~~~~~~~~~~~~~~~~~~~~~~~~~~~~~~~~~~~~~~~~~~~~~~~~~~~~~~$$
\begin{equation}\label{20}
~~~~~~~~~~~~~~~~~~~~~~~~~~~~~~~~~~~~~~~~~~~~~~~~~~~~~~~~~~~~~~~~~~~~~+\frac{4 a_{vw} b_{vw} \pi \log{\left(\frac{3}{2}+\frac{r_c}{b_{vw}}\right)}}{r^2_c} > \frac{3b_{vw}}{2l^2}+\frac{4a_{vw} \pi}{r_c\left(\frac{r_c}{b_{vw}}+\frac{3}{2}\right)}
\end{equation}
and
$$ 2\pi a_{vw}+ \frac{r_c^2}{l^2}\left(1+\frac{3b_{vw}}{2r_c}\right)+\frac{r_c}{4} \Bigg[ \frac{2M}{r^2_c}+\frac{2\left(1+\frac{3b_{vw}}{2r_c}\right)r_c}{l^2}+ \frac{6 a_{vw} b_{vw}^2 \pi}{r_c(2r_c+3b_{vw})^2}+\frac{3\pi a_{vw} b_{vw}^2}{r_c^2(2r_c+3b_{vw})}~~~~~~~~~~~~~~~~~~~~~~~~~~~~~~~~~~~~~~~~~$$
\begin{equation}
+\frac{4 a_{vw} b_{vw} \pi \log{\left(\frac{3}{2}+\frac{r_c}{b_{vw}}\right)}}{r^2_c}\Bigg] > \frac{r_c}{4}\Bigg[ \frac{3b_{vw}}{2l^2}+\frac{4a_{vw} \pi}{r_c\left(\frac{r_c}{b_{vw}}+\frac{3}{2}\right)}\Bigg] + \frac{2M}{r_c} +\frac{3\pi a_{vw} b_{vw}^2}{r_c(2r_c+3b_{vw})} + \frac{4\pi a_{vw} b_{vw}}{r_c}\log{\left(\frac{r_c}{b_{vw}}+\frac{3}{2}\right)}. 
\end{equation}
Now, we consider $p=A \rho$ is the equation of state and $A$ is constant and it accretes upon the Van der Waal's black hole. Then we get $c_s^2=A, ~V_c^2=0$ and $u_c^2=0$.

Therefore, from (\ref{12}), we get,
\begin{equation}\label{23}
\frac{2M}{r_c^2}+\frac{2(1+\frac{3b_{vw}}{2r_c})r_c}{l^2}+\frac{6a_{vw}b_{vw}^2 \pi}{r_c(3b_{vw}+2r_c)^2}+\frac{3a_{vw}b_{vw}^2 \pi}{r_c^2(3b_{vw}+2r_c)}+\frac{4a_{vw}b_{vw} \pi log(\frac{3}{2}+\frac{r_c}{b_{vw}})}{r_c^2}= \frac{3b_{vw}}{2l^2}+\frac{4a_{vw} \pi}{r_c(\frac{3}{2}+\frac{r_c}{b_{vw}})}
\end{equation}
Let $\dot{{M}}$ is the rate of change of mass of Van der Waal's black hole which is computed by integrating the flux over the two dimensional surface of the black hole and is defined by\cite{John1}, 
\begin{equation}\label{26}
\dot{{M}}=- 4 \pi r^2 T_0^1~~\Rightarrow~~\dot{{M}} = 4 \pi {\cal C}\{\rho_\infty + p(\rho_\infty)\}
\end{equation}
If we assume $M_0$ be the initial mass corresponding to the initial time and if we neglect the cosmological evolution of $\rho_\infty$, then using the equation (\ref{26}), we get the mass of the black hole to be
\begin{equation}\label{28}
M = M_0 + 4 \pi {\cal C}\{\rho_\infty + p(\rho_\infty)\}(t-t_0)
\end{equation}
The result (\ref{26}) can be written for any general $\rho$ and $p$ as done in \cite{Jamil2, Jamil3, Jamil4} (satisfying the holographic equation of state and violating weak energy condition), i.e., can be written as   
\begin{equation}\label{29}
\dot{{M}} = 4 \pi {\cal C} (\rho+p)
\end{equation}
Again, black hole simultaneously radiates energy when it accretes fluid . This radiation is known as Hawking radiation\cite{Susskind1}. The black hole evaporates for this radiation which is balanced by the accretion of matter into the black hole and as a outcome the total system is guessed to be under equilibrium. But when we examine the parameters (eg., temperature) of the accreting fluid at very far from the black hole with very close to the black hole, there will be a big difference. But the parameters show equilibrium nature in local cells. Such type of equilibrium is called quasi equilibrium. In this work, we have considered large black holes (in general). For small black holes, the relation between temperature and mass is given by $T = (8 \pi M)^{-1}$. For this reason, the black holes radiate more following to the standard fourth order rule of black body radiation. The accretion radiation equilibrium may not be the equilibrium one under such large amount of Hawking radiation. For this type of cases we will unable to talk whether the accretion process is at all dependent of the mass or not. The process of accretion for very small black holes is still a fact to research with. From the equation (\ref{s2}) and (\ref{7}), we have (with $r_h = 1$) the index of the equation of state as 
\begin{equation}\label{30}
\omega_D = -1 + {\cal C}_4 \Bigg\{u^2 + 2\pi a_{vw}- \frac{2M}{r} + \frac{r^2}{l^2}\left(1+\frac{3b_{vw}}{2r}\right)-\frac{3\pi a_{vw} b_{vw}^2}{r(2r+3b_{vw})} - \frac{4\pi a_{vw} b_{vw}}{r}\log{\left(\frac{r}{b_{vw}}+\frac{3}{2}\right)} \Bigg\}^{-1/2}
\end{equation} 
Note: $\omega_D > or < -1~~depends~on~the~sign~of~the~constant~{\cal C}_4.$
\section{Thermodynamic Analysis of Accreting Matter on Van der Waal's Black Hole}
%%%%%%%%%%%%%%%%%%%%%%%%%%%%%%%%%%%%%%%%%%%%%%%%%%%%%%%%%%%%%%%%%%%%%%%%
Now, we will discuss about the thermodynamics of the dark energy accretion. The equation related to the thermodynamic studies is given by the equation of state $p = \omega_D \rho$. First, we wish to evaluate the value of ${\cal C}$ such that the sign of $M$ can be determined and secondly, we verify the exactness of the generalized second law of thermodynamics which is an invariant law and will search any limitation on the equation of state $\omega_D$ from thermodynamic point of view. The energy supply vector $\psi_i$ and the work density($W$) are defined as\cite{Cai2, Chakraborty1}

$W = - \frac{1}{2} Trace \{T^i_j\} = \frac{1}{2}(\rho - 3p)$ and $\psi_i = T^i_j \partial_j r + W \partial_i r$,

i.e., $\psi_0 = T^1_0 = -u(\rho+p)\sqrt{u^2 + 2\pi a_{vw}- \frac{2M}{r} + \frac{r^2}{l^2}\left(1+\frac{3b_{vw}}{2r}\right)-\frac{3\pi a_{vw} b_{vw}^2}{r(2r+3b_{vw})} - \frac{4\pi a_{vw} b_{vw}}{r}\log{\left(\frac{r}{b_{vw}}+\frac{3}{2}\right)} }$ 

and $\psi_1 = T_1^1 + W = \rho \left\{\frac{1}{2} + \frac{u^2}{2\pi a_{vw}- \frac{2M}{r} + \frac{r^2}{l^2}\left(1+\frac{3b_{vw}}{2r}\right)-\frac{3\pi a_{vw} b_{vw}^2}{r(2r+3b_{vw})} - \frac{4\pi a_{vw} b_{vw}}{r}\log{\left(\frac{r}{b_{vw}}+\frac{3}{2}\right)} }\right\} $

$~~~~~~~~~~~~~~~~~~~~~~~~~~~~~~~~~~~~~~~~~~~~~~~~~~~~~~~~~~~~~~~~~~~~~~~~~~~~+ p\left\{-\frac{1}{2} + \frac{u^2}{2\pi a_{vw}- \frac{2M}{r} + \frac{r^2}{l^2}\left(1+\frac{3b_{vw}}{2r}\right)-\frac{3\pi a_{vw} b_{vw}^2}{r(2r+3b_{vw})} - \frac{4\pi a_{vw} b_{vw}}{r}\log{\left(\frac{r}{b_{vw}}+\frac{3}{2}\right)} }\right\}$, 

where $T_i^j$ is the projected energy-momentum tensor (normal to the 2-sphere). Therefore, the change of energy (across the event horizon) is given by\cite{Chakraborty1}
$$-dE = -A\psi = -A[\psi_0 dt + \psi_1 dr ]$$
The amount of energy crossing the event horizon is\cite{Chakraborty1, Mazumder1} (taking $r_e = 1$) given by
\begin{equation}\label{31}
dE = 4\pi u^2 (\rho + p) dt.
\end{equation}
From (\ref{26}) and (\ref{31}) (as $c=1$ and $E=mc^2$), we get, the arbitrary constant ${\cal C}$, given by

$~~~~~~~~~~~~~~~~~~~~~~~~~~~~~~~~~~~~~~~~~~~~~~~~~~~~~~{\cal C}=u^2$, i.e., $\dot{M}= 4\pi u^2 (\rho + p)~~.$

In quintessence era, we can say that $\dot{M} > 0$, i.e., the black hole mass is increasing although the rate of increment is slowly decreasing as we move to the line of phantom barrier, whereas, in phantom era, $\dot{M} < 0$, i.e., the black hole mass is decreasing. The holographic energy density given by 
\begin{equation}\label{32}
\rho = \frac{3c^2}{R_h^2}~~~~,
\end{equation}   
where $R_h^2 = a \int_t^{\infty} \frac{dt}{a} = a \int_t^{\infty} \frac{da}{H a^2}$ which directs to result balanced with observations. Here `$a$' is the scale factor of the background metric of the universe and $H$ is the Hubble parameter.

We can identify the dimensionless dark energy density parameter as:
\begin{equation}\label{33}
\Omega_h = \frac{\rho}{3 H^2} = \frac{c^2}{R_h^2 H^2}
\end{equation}
For a dark energy subjected universe, dark energy enlarge similar to the conservation law
\begin{equation}\label{34}
\dot{\rho} + 3H(\rho + p) = 0
\end{equation}
or identically\cite{Biswas1}:
\begin{equation}\label{35}
\dot{\Omega_h} = H \Omega_h(1 - \Omega_h)\left(1 + 2 \frac{\sqrt{\Omega_h}}{c}\right)~~~~~,
\end{equation}
where $p=\omega_D \rho$ is the equation of state.

Also, the equation of state of the index is of the form\cite{Biswas1}:
\begin{equation}\label{36}
\omega_D = - \frac{1}{3}\left(1 + 2 \frac{\sqrt{\Omega_h}}{c}\right).
\end{equation}
Here, $w_D$ depends on the parameter $c$. Since, the observation predicts\cite{Huang1} $\Omega_h \rightarrow 1$ for the present time, therefore, at $c=1$, $\omega_D \rightarrow -1$, i.e., our model acts like cosmological constant. Also for $c>1$, we get, $-1<\omega_D<-\frac{1}{3}$, i.e., our model shows the quintessence region and if $c<1$, we get, $\omega_D<-1$, i.e., the phantom type behaviour occurs. 

Using the equation (\ref{32}) and (\ref{36}), we get,
\begin{equation}\label{37}
\dot{M} = 8 \pi u^2 \frac{c^2}{R_h^2}\left(1 - \frac{1}{R_h H}\right)\Rightarrow \frac{d \dot{M}}{dR_h} = 8 \pi u^2 \frac{c^2}{R_h^4}\left(\frac{3}{H} - 2 R_h\right)~~.
\end{equation} 
If $R_h < \frac{3}{2}R_H$, then $\dot{M}$ increases where $R_H$ is the Hubble radius and $R_h$ is the radius of the event horizon. If $R_h > \frac{3}{2}R_H$, then $\dot{M}$ decreases.

\section{Dark Energy Accretion upon Van der Waal's Black Hole}

Here, we will discuss about dark energy model such as extended Chaplygin gas. We consider the spatially flat, homogeneous and isotropic Friedmann-Robertson-Walker (FRW) model of the universe is described by the following metric
$$ds^2 =-dt^2+a^2(t) \left[ dr^2 +r^2d\Omega^2 \right]~~~~~~,$$
where $a(t)$ represents time-dependent scale factor.

The Einstein's equations for FRW universe are 
\begin{equation}\label{39}
H^2 =\frac{1}{3} \rho~~and
\end{equation}
\begin{equation}\label{40}
\dot{H}=-\frac{1}{2}(p+ \rho)~~.
\end{equation}
It is also assumed that the total matter and energy are conserved with the following conservation equation (\ref{34})

Now, we consider the extended Chaplygin gas\cite{Debnath1} as dark energy model. The equation of state is given by
\begin{equation}\label{42}
p=\sum_n~A_n \rho^n -\frac{B}{\rho^{\alpha}}
\end{equation}
{\bf ~~~~~I. For }$n=1$

Special case of $n=1$ reduces the equation (\ref{42}) to the modified Chaplygin gas equation of state with the density (using the equation(\ref{34}))
\begin{equation}\label{43}
\rho = \left[ \frac{B}{1+A} + \frac{C}{a^{3(1+\alpha)(1+A)}} \right]^{\frac{1}{(1+\alpha)}}~~~,
\end{equation}
where $C>0$ is an integration constant.

Therefore, the current value of the energy density 
\begin{equation}\label{44}
\rho_0=\left[ \frac{B}{1+A} +C \right]^{\frac{1}{(1+\alpha)}} ~~.
\end{equation}
For MCG model, we get,
\begin{equation}
c_s^2=A+\frac{\alpha B}{\rho^{1+\alpha}} 
\end{equation}
and 
$$V_c^2=\frac{(1+\alpha) B}{(1+A)\rho^{1+\alpha}-B}$$
$$\Rightarrow V_c^2= \Bigg[\frac{B}{1+A}\Bigg\{1+(1+\alpha)\Bigg(1+\frac{4}{r_c}~~~~~~~~~~~~~~~~~~~~~~~~~~~~~~~~~~~~~~~~~~~~~~~~~~~~~~~~~~~~~~~~~~~~~~~~~~~~~~~~~~~~~~~~~~~~~~~~~~~~~~~~~~~~~$$
\begin{equation}
~~~~~~~~~~~~~~~\times\frac{ 2\pi a_{vw}- \frac{2M}{r_c} + \frac{r_c^2}{l^2}\left(1+\frac{3b_{vw}}{2r_c}\right)-\frac{3\pi a_{vw} b_{vw}^2}{r_c(2r_c+3b_{vw})} - \frac{4\pi a_{vw} b_{vw}}{r_c}\log{\left(\frac{r_c}{b_{vw}}+\frac{3}{2}\right)}}{-\frac{3b_{vw}}{2l^2}+\frac{2M}{r^2_c}+\frac{2\left(1+\frac{3b_{vw}}{2r_c}\right)r_c}{l^2}+ \frac{6 a_{vw} b_{vw}^2 \pi}{r_c(2r_c+3b_{vw})^2}+\frac{3\pi a_{vw} b_{vw}^2}{r_c^2(2r_c+3b_{vw})}-\frac{4a_{vw} \pi}{r_c\left(\frac{r_c}{b_{vw}}+\frac{3}{2}\right)}+\frac{4 a_{vw} b_{vw} \pi \log{\left(\frac{3}{2}+\frac{r_c}{b_{vw}}\right)}}{r^2_c}}\Bigg)\Bigg\}\Bigg]^{\frac{1}{1+\alpha}}~~~~~~~~~~~~~~~~~~~~~~~~~~~~~~~~~~~~~~~~~~~~~~~~~~~~~~~~~~~~~~~~~~~~~~~
\end{equation}
In fig.1a we have plotted $\rho$ vs $M$, the mass of the central gravitating object. Accretion has been considered. The plots show  if the mass of the central engine is increased, the density of the accreting dark energy is reduced. Not only that but also we observe that the density profile is high if $\alpha$ is low. So dark energy, whenever is strongly repulsive (i.e., $\alpha$ is high) we find it to reduce the density to be accreted in. Super massive black holes are less capable to accrete a strongly dense dark energy flow towards it than the local stellar mass black holes can do.  For $\alpha=0$, we get back the barotropic fluid accretion. We see for small mass of the black hole, the accretion is very high (than dark energy cases) and as black hole's mass is increased, density of the exotic fluid decreases. The negative pressure of dark energy faints the strength of accretion.
\begin{figure}[ht]
\begin{center}

~~~~~~~~~~~~~~~~~~~~~~~~~~~~Fig.1a~~~~~~~~~~~~~~~~~~~~~~~~~~~~~~~~~~~~~~~~~\\
\vspace{0.5cm}
\includegraphics[height=2.5in, width=3.2in]{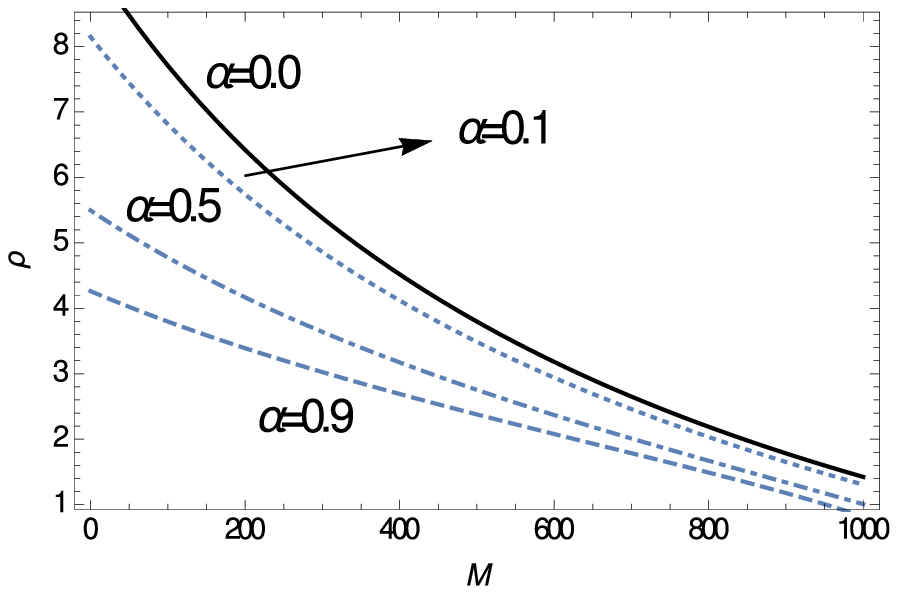}\\
\vspace{0.1cm}
Relation between $\rho$ and $M$ 
$$r_c=10, a_{vw}=0.2, b_{vw}=1, l=1.2, A=1/3, B=3$$
\end{center}
\end{figure}

{\bf II. For }$\alpha=-1$

Here, we assume the last term of expression in EoS (\ref{42}) is dominant. In this case, we can write the energy density in terms of the scale factor as (using the equation(\ref{34}))
\begin{equation}\label{45}
\rho = \left[\frac{A}{B-1} + \frac{C}{a^{3(B-1)(n-1)}} \right]^{\frac{1}{1-n}}~~~,
\end{equation}
where $C$ is an arbitrary integrating constant.

Therefore, the current value of the energy density looks like
\begin{equation}\label{46}
\rho_0 = \left[\frac{A}{B-1} + C \right]^{\frac{1}{1-n}}
\end{equation}
Also, we get,
\begin{equation}
c_s^2= \sum_{i=1}^n~iA_i \rho^{i-1} - B
\end{equation} 
and 
$$V_c^2=\frac{\sum_{i=1}^{n-1}~iA_{i+1} \rho^i}{1+\sum_{i=1}^n~A_i \rho^{i-1}-B}$$
\begin{equation}\Rightarrow V_c^2= 1+\frac{4}{r_c}\frac{ 2\pi a_{vw}- \frac{2M}{r_c} + \frac{r_c^2}{l^2}\left(1+\frac{3b_{vw}}{2r_c}\right)-\frac{3\pi a_{vw} b_{vw}^2}{r_c(2r_c+3b_{vw})} - \frac{4\pi a_{vw} b_{vw}}{r_c}\log{\left(\frac{r_c}{b_{vw}}+\frac{3}{2}\right)}}{-\frac{3b_{vw}}{2l^2}+\frac{2M}{r^2_c}+\frac{2\left(1+\frac{3b_{vw}}{2r_c}\right)r_c}{l^2}+ \frac{6 a_{vw} b_{vw}^2 \pi}{r_c(2r_c+3b_{vw})^2}+\frac{3\pi a_{vw} b_{vw}^2}{r_c^2(2r_c+3b_{vw})}-\frac{4a_{vw} \pi}{r_c\left(\frac{r_c}{b_{vw}}+\frac{3}{2}\right)}+\frac{4 a_{vw} b_{vw} \pi \log{\left(\frac{3}{2}+\frac{r_c}{b_{vw}}\right)}}{r^2_c}}
\end{equation}
Density vs mass for $\alpha=-1$ case has been plotted in fig.1b. The basic features do match with fig.1b. However, as we increase the value of $n$, the initial (for low $M$) decrease of density becomes more steeper. If we take $n=1$, it is obsereved that the accretion density almost becomes constant with the variation of mass $M$.
\begin{figure}[ht]
\begin{center}

~~~~~~~~~~~~~~~~~~~~~~~~~~~~~~~~~~~~~Fig.1b~~~~~~~~~~~~~~~~~~~~~~~~~~~~~~~~~~~~~~~~~\\
\vspace{0.5cm}
\includegraphics[height=2.5in, width=3.2in]{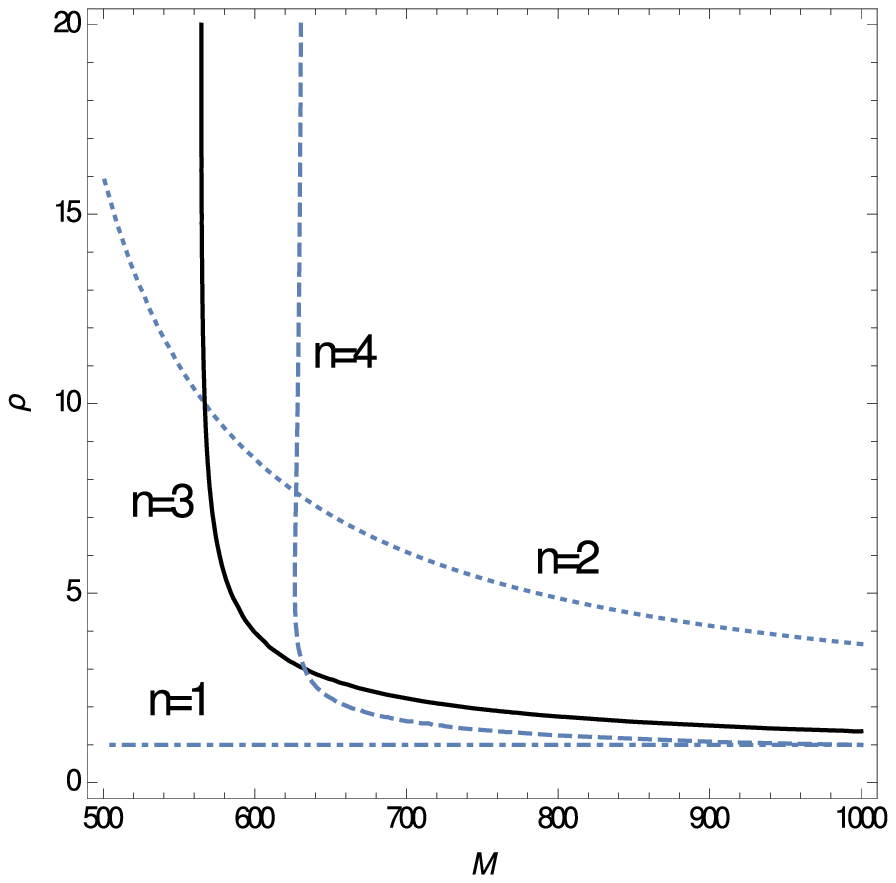}\\
\vspace{0.1cm}
Relation between $\rho$ and $M$

$r_c=10,~a_{vw}=0.2,~b_{vw}=1,~l=1.2,~A=A_1=A_2=A_3=A_4=\frac{1}{3},~B=3$ 

\end{center}
\end{figure}

{\bf III. For }$n=2$ and $\alpha=\frac{1}{2}$

In this case the equation of  state given by equation(\ref{42}) can be written as,
\begin{equation}\label{47}
p=A_1 \rho+ A_2 \rho^2 - \frac{B}{\sqrt{\rho}}~~.
\end{equation}
Using the equation (\ref{34}), we get,
\begin{equation}\label{48}
ln(a) = - \int \frac{d\rho}{3\left\{(1+A_1)\rho+A_2\rho^2-\frac{B}{\sqrt{\rho}}\right\}} ~~.
\end{equation}
Therefore, the energy density is given by
\begin{equation}\label{49}
\rho = \phi_1(a)~~~,
\end{equation}
where $\phi_1$ can be evaluated by expressing $\rho$ as a function of $a$ by using the equation (\ref{48}). 

Therefore, the current value of the energy density is $\rho_0 = \phi_1(1)$.

Also, we get,
\begin{equation}
c_s^2 = A_1 + 2A_2 \rho + \frac{B}{2\rho^{3/2}}
\end{equation}  
and
$$V_c^2 = \frac{A_2 \rho+\frac{3}{2}\frac{B}{\rho^{3/2}}}{1+A_1+A_2 \rho- \frac{B}{\rho^{3/2}}}$$
\begin{equation}
\Rightarrow V_c^2= 1+ \frac{4}{r_c}\frac{2\pi a_{vw}- \frac{2M}{r_c} + \frac{r_c^2}{l^2}\left(1+\frac{3b_{vw}}{2r_c}\right)-\frac{3\pi a_{vw} b_{vw}^2}{r_c(2r_c+3b_{vw})} - \frac{4\pi a_{vw} b_{vw}}{r_c}\log{\left(\frac{r_c}{b_{vw}}+\frac{3}{2}\right)}}{-\frac{3b_{vw}}{2l^2}+\frac{2M}{r^2_c}+\frac{2\left(1+\frac{3b_{vw}}{2r_c}\right)r_c}{l^2}+ \frac{6 a_{vw} b_{vw}^2 \pi}{r_c(2r_c+3b_{vw})^2}+\frac{3\pi a_{vw} b_{vw}^2}{r_c^2(2r_c+3b_{vw})}-\frac{4a_{vw} \pi}{r_c\left(\frac{r_c}{b_{vw}}+\frac{3}{2}\right)}+\frac{4 a_{vw} b_{vw} \pi \log{\left(\frac{3}{2}+\frac{r_c}{b_{vw}}\right)}}{r^2_c}}
\end{equation}
In fig.1c we have plotted $\rho$ vs $M$ and we fix $r_c,~a_{vw},~b_{vw},~l,~A_1,~A_2$ and $~B$, and see that if the mass is increasing then the density is decreasing.  
\begin{figure}[ht]
\begin{center}

~~~~~~~~~~~~~~~~~~~~~~~~~~~~~~~~~~~~~Fig.1c~~~~~~~~~~~~~~~~~~~~~~~~~~~~~~~~~~~~~~~~~\\
\vspace{0.5cm}
\includegraphics[height=2.5in, width=3.2in]{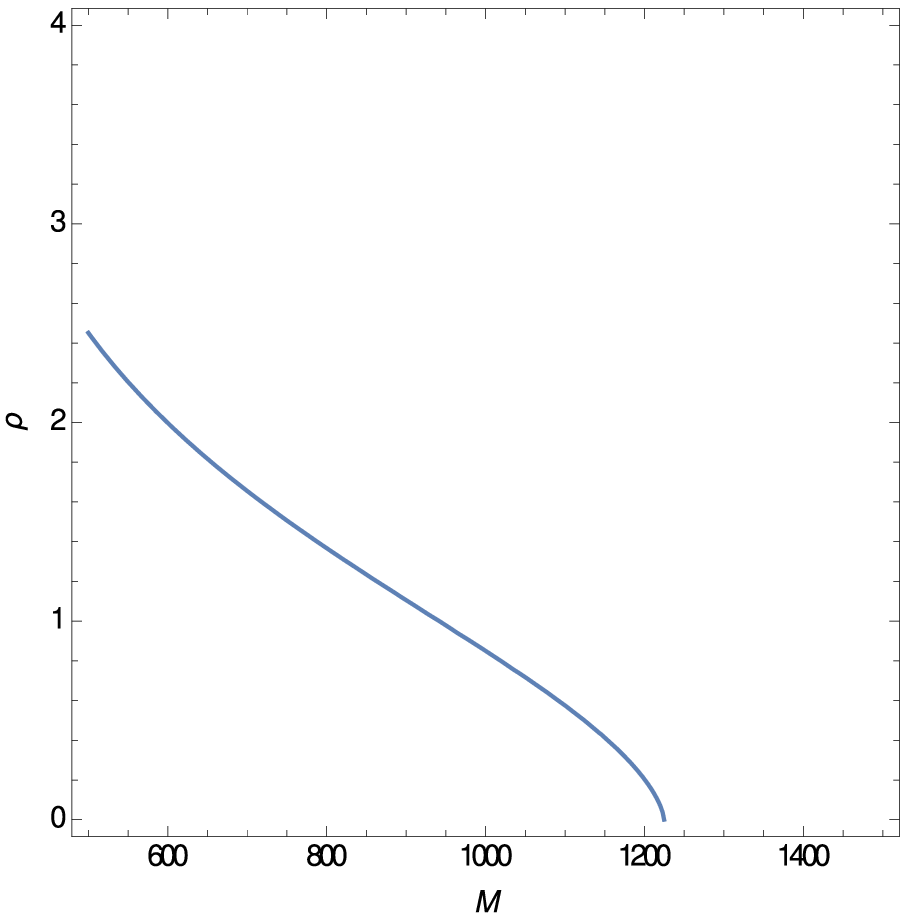}\\
\vspace{0.1cm}
Relation between $\rho$ and $M$

$r_c=10,~a_{vw}=0.2,~b_{vw}=1,~l=1.2,~A_1=A_2=\frac{1}{3},~B=3$ 

\end{center}
\end{figure}

{\bf IV. For $n=3$ and }$\alpha=\frac{1}{2}$

In this case the EoS (\ref{42}) can be written as,
\begin{equation}\label{51}
p=A_1 \rho+ A_2 \rho^2 +A_3 \rho^3 - \frac{B}{\sqrt{\rho}}
\end{equation}
Using the equation (\ref{34}), we get,
\begin{equation}\label{52}
ln(a) = - \int \frac{d\rho}{3\left\{(1+A_1)\rho+A_2\rho^2+A_3 \rho^3-\frac{B}{\sqrt{\rho}}\right\}} 
\end{equation}

Therefore, the current value of the energy density $\rho_0 = \phi_2(1)$.

Also, we get,
\begin{equation}
c_s^2 = A_1 + 2A_2 \rho+3A_3 \rho^2  + \frac{B}{2\rho^{3/2}}
\end{equation}  
and 
$$V_c^2 = \frac{A_2 \rho+2A_3 \rho^2+\frac{3}{2}\frac{B}{\rho^{3/2}}}{1+A_1+A_2 \rho+A_3 \rho^2- \frac{B}{\rho^{3/2}}}$$
\begin{equation}
\Rightarrow V_c^2= 1+\frac{4}{r_c}\frac{2\pi a_{vw}- \frac{2M}{r_c} + \frac{r_c^2}{l^2}\left(1+\frac{3b_{vw}}{2r_c}\right)-\frac{3\pi a_{vw} b_{vw}^2}{r_c(2r_c+3b_{vw})} - \frac{4\pi a_{vw} b_{vw}}{r_c}\log{\left(\frac{r_c}{b_{vw}}+\frac{3}{2}\right)}}{-\frac{3b_{vw}}{2l^2}+\frac{2M}{r^2_c}+\frac{2\left(1+\frac{3b_{vw}}{2r_c}\right)r_c}{l^2}+ \frac{6 a_{vw} b_{vw}^2 \pi}{r_c(2r_c+3b_{vw})^2}+\frac{3\pi a_{vw} b_{vw}^2}{r_c^2(2r_c+3b_{vw})}-\frac{4a_{vw} \pi}{r_c\left(\frac{r_c}{b_{vw}}+\frac{3}{2}\right)}+\frac{4 a_{vw} b_{vw} \pi \log{\left(\frac{3}{2}+\frac{r_c}{b_{vw}}\right)}}{r^2_c}} 
\end{equation}
\begin{figure}[ht]
\begin{center}

~~~~~~~~~~~~~~~~~~~~~~~~~~~~~~~~~~~~~Fig.1d~~~~~~~~~~~~~~~~~~~~~~~~~~~~~~~~~~~~~~~~~\\
\vspace{0.5cm}
\includegraphics[height=2.0in, width=3.0in]{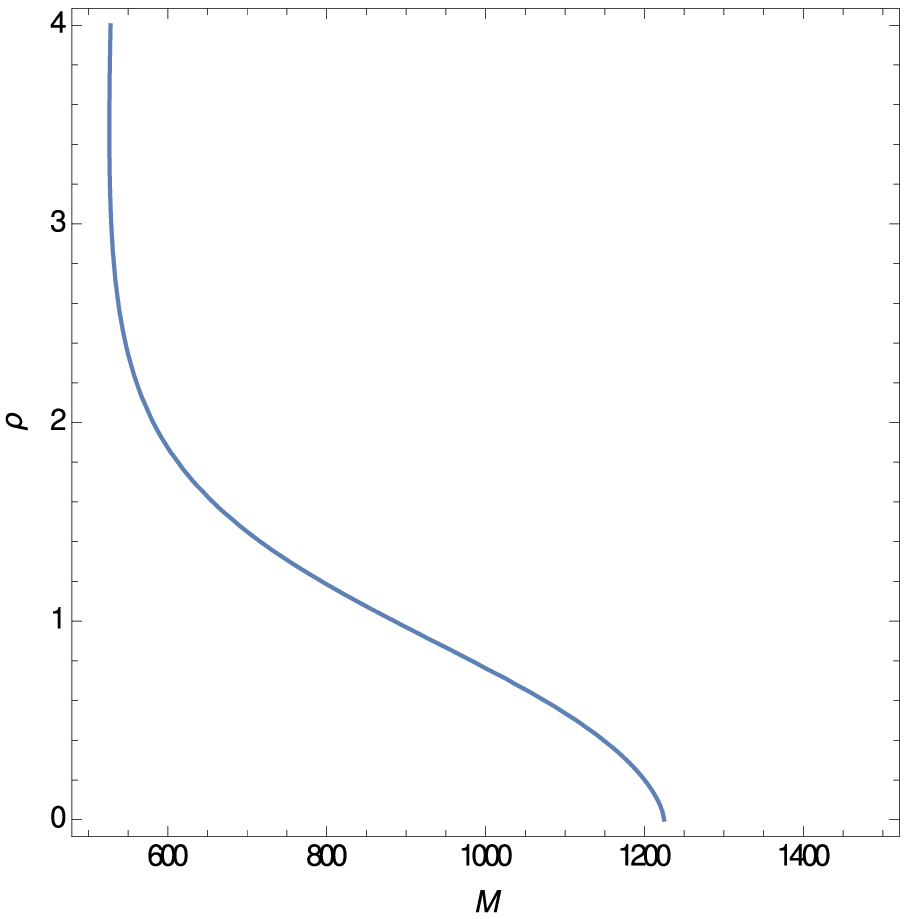}\\
\vspace{0.1cm}
Relation between $\rho$ and $M$

$r_c=10,~a_{vw}=0.2,~b_{vw}=1,~l=1.2,~A_1=A_2=A_3=\frac{1}{3},~B=3$ 

\end{center}
\end{figure}

Density vs mass for ``$n=3$ and $\alpha=\frac{1}{2}$" case has been plotted in fig. 1d. The basic features do match with figures 1a, 1b and 1c.   

Now, using the equation (\ref{34}), (\ref{39}) and (\ref{40}), we get,
\begin{equation}
\dot{M}=-\frac{4 \pi u^2}{\sqrt{3}} \frac{\dot{\rho}}{\sqrt{\rho}}~~~~,
\end{equation}
which implies
\begin{equation}\label{56}
M=M_0- \frac{8 \pi u^2}{\sqrt{3}}(\sqrt{\rho}-\sqrt{\rho_0})~~~,
\end{equation}
where $M_0$ is the current value of the Van der Waal's black hole's mass. If $a$ is very large ($z\rightarrow -1$), i.e., at the last stage of the universe, the mass of the black hole will be $M=M_0 + \frac{8 \pi u^2}{\sqrt{3}}\sqrt{\rho_0}$ .

Using the solution of $\rho$ in equation (\ref{56}), the black hole mass $M$ can be written in terms of scale factor $a$ and then using $z=\frac{1}{a}-1$, the formula of redshift, $M$ will be written in terms of redshift $z$.

For $n=1$, $M$ can be written as, 
\begin{equation}
M=M_0+\frac{8 \pi u^2}{\sqrt{3}}\left[\left(C+\frac{B}{1+A}\right)^{\frac{1}{2(1+\alpha)}}-\left\{C(1+z)^{3(1+A)(1+\alpha)}+\frac{B}{1+A}\right\}^{\frac{1}{2(1+\alpha)}}\right]
\end{equation}
\begin{figure}[ht]
\begin{center}
~~~~~~~~~~~~~~~~~~~~~~~~~~~~~~~~~~~~~Fig.2a~~~~~~~~~~~~~~~~~~~~~~~~~~~~~~~~~~~~~~~~~\\
\vspace{0.5cm}
\includegraphics[height=2.0in, width=3.5in]{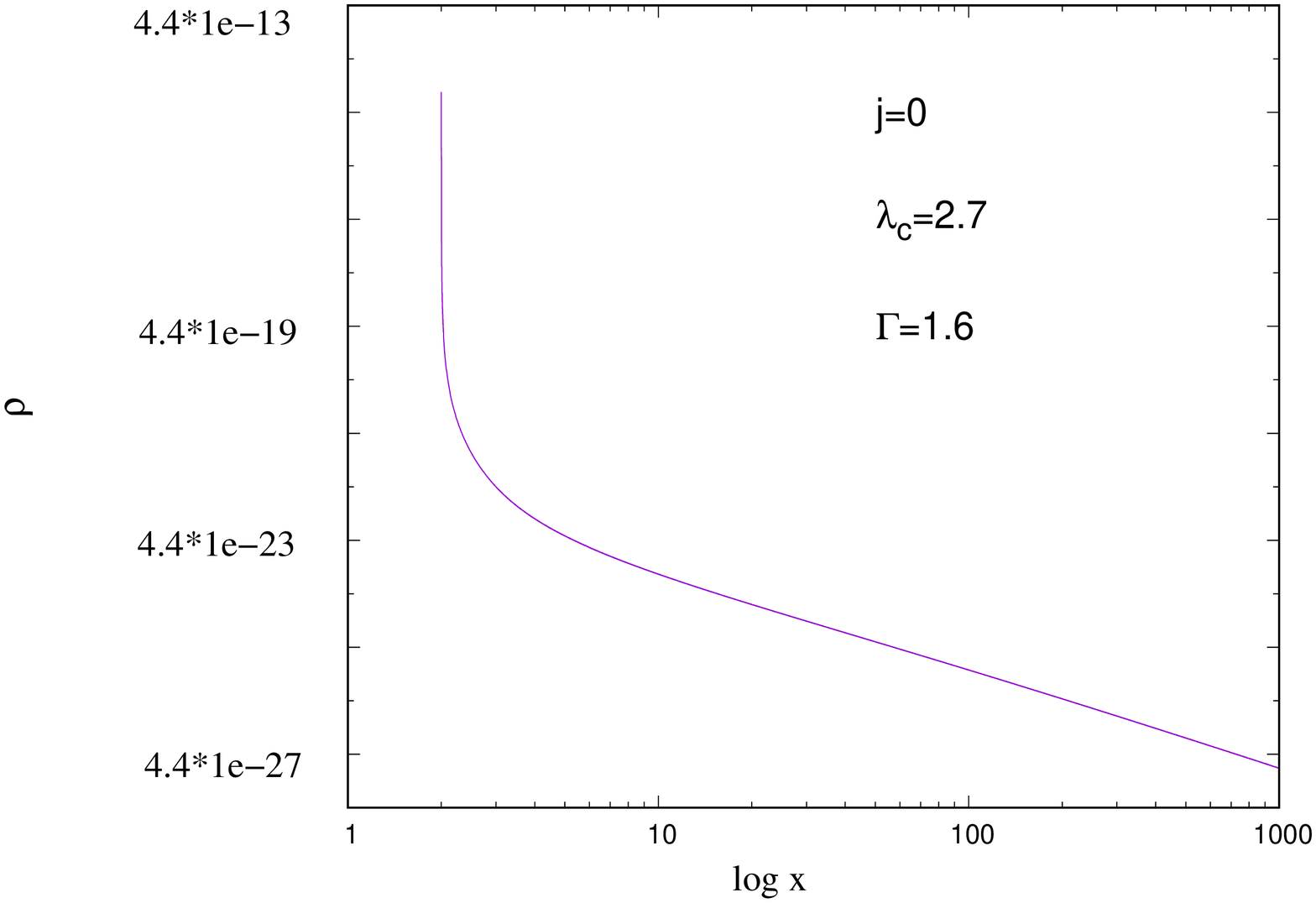}\\
\vspace{0.1cm}
Relation between $\rho$ and $x$ for adiabatic accretion on Schwarzschild black hole.   
\end{center}
\end{figure}

We will now compare Chaplygin gas accretion for Van der Waal's black hole and Schwarzschild black hole. A quantitative study of Chaplygin gas accretion and detailed density profile are studied in the reference \cite{Sandip2}. For a nonrotating ($j=0$) i.e., Schwarzschild black hole, we observe that if the specific angular momentum ($\lambda_c$) of the accretion is $2.7$, for adiabatic accretion (Fig 2a) the density of accretion near to the black hole is very high ($\sim 10^{-13}$ gm/cc), whereas, at thousand Schwarzschild radius this will be of the order of$10^{-27} gm/cc$. As we increase $x$, we observe that the density become asymptotic to the distance axis.

If we observe the properties of Van der Waals's fluid besides the ideal fluid, we see that the Van der Waal's fluid incorporates the volume occupied by the molecules and the increment in pressure due to their collisions. This is analogical to the effects created by the microsates of the black holes. It is expected that the existence of different microstates will reduce the ``as a whole accreting power" of a black hole (due to their mutual interactions). We observe in figures 1a-1d that as we introduce the effects of Van der Waal's black hole very prominently, the accreting density reduces after a particular measurement of black hole mass M ( distance =$x \frac{GM}{c^2}$). This means the Van der Waal's parameters increase the mutual interactions between the microstates which result reduction in accreting power. This makes the accretion disk of a finite length.

\begin{figure}[ht]
\begin{center}
~~~~~~~~~~~~~~~~~~~~~~~~~~~~~~~~~~~~~Fig.2b~~~~~~~~~~~~~~~~~~~~~~~~~~~~~~~~~~~~~~~~~\\
\vspace{0.5cm}
\includegraphics[height=2.0in, width=3.5in]{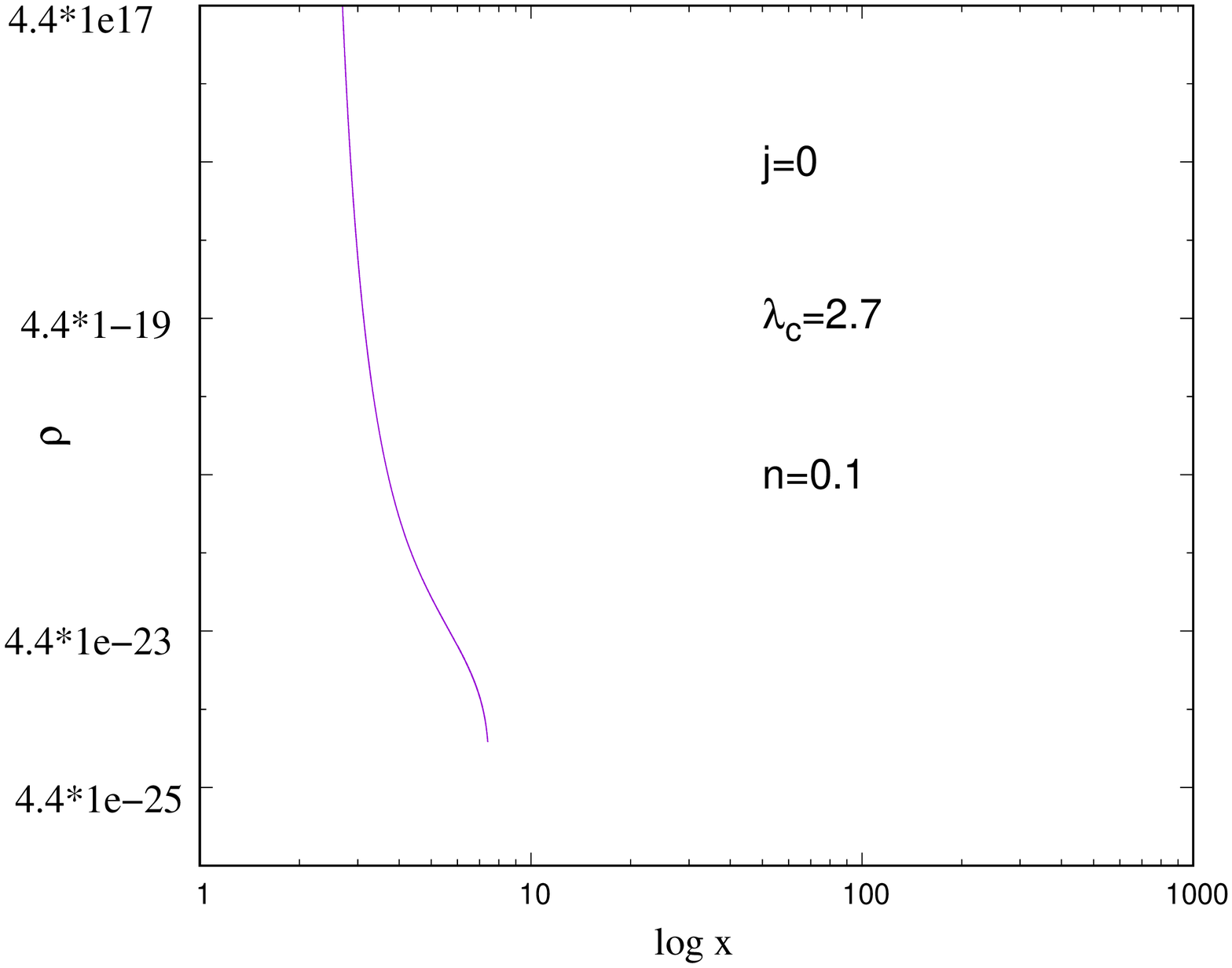}\\
\vspace{0.1cm}
Relation between $\rho$ and $x$ for Chaplygin gas accretion on Schwarzschild black hole. 

\end{center}
\end{figure}

Finally, in fig 2b, we draw the Chaplygin gas accretion on Schwarzschild black hole \cite{Sandip2}. Here, we see effect of dark energy accretion terminates the accretion disc very near to the black hole.

So we can speculate that the Chaplygin gas type dark energy accretion on the Van der Waal's black hole is stronger than the dark energy accretion on Schwarzschild black hole (i.e., the Van der Waal's black hole is able to even attract dark energy from distant regions) and is weaker than an adiabatic fluid accretion on Schwarzschild black hole (i.e., Schwarzschild black hole is able to accrete adiabatic fluid from distant region but Van der Waal's black hole is unable do that).  

\begin{figure}[ht]
\begin{center}

~~~~~~~~~~~~~~~~~~~~~~~~~~~~Fig.3a(i)~~~~~~~~~~~~~~~~~~~~~~~~~~~~~~~~~~~~~~~~~\\
\vspace{0.5cm}
\includegraphics[height=2.0in, width=3.0in]{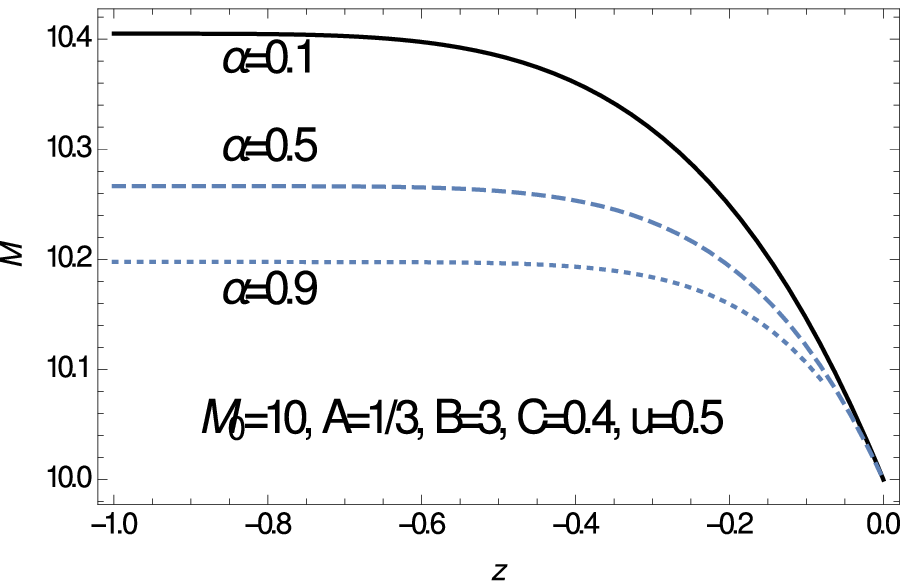}\\
\vspace{0.1cm}
Relation between $M$ and $z$ 
$$M_0=10, A=1/3, B=3, C=0.4, u=0.5$$
\end{center}
\end{figure}

\begin{figure}[ht]
\begin{center}

~~~~~~~~~~~~~~~~~~~~~~~~~~~~Fig.3a(ii)~~~~~~~~~~~~~~~~~~~~~~~~~~~~~~~~~~~~~~~~~\\
\vspace{0.5cm}
\includegraphics[height=2.0in, width=3.0in]{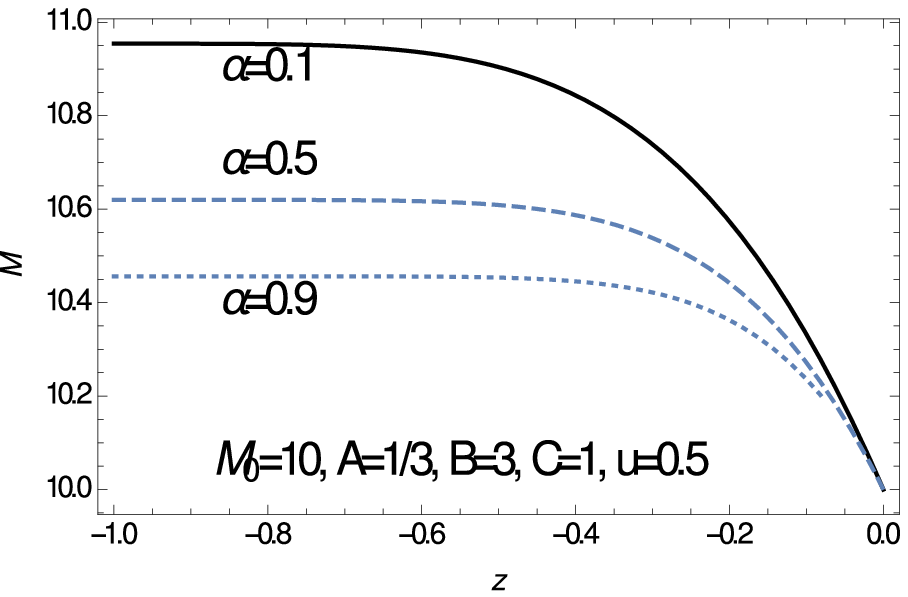}\\
\vspace{0.1cm}
Relation between $M$ and $z$ 
$$M_0=10, A=1/3, B=3, C=1, u=0.5$$
\end{center}
\end{figure}
\begin{figure}[ht]
\begin{center}

~~~~~~~~~~~~~~~~~~~~~~~~~~~~Fig.3a(iii)~~~~~~~~~~~~~~~~~~~~~~~~~~~~~~~~~~~~~~~~~\\
\vspace{0.5cm}
\includegraphics[height=2.0in, width=3.0in]{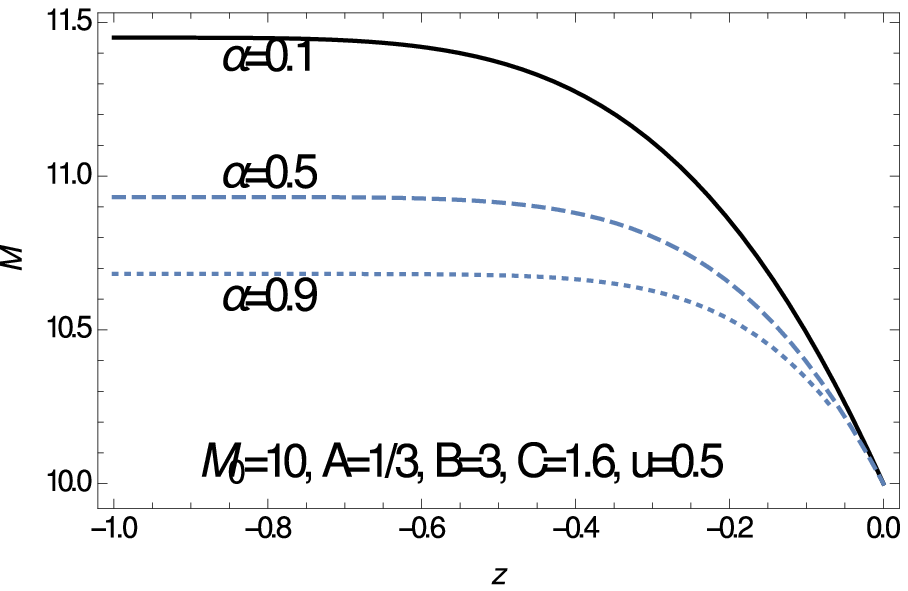}\\
\vspace{0.1cm}
Relation between $M$ and $z$ 
$$M_0=10, A=1/3, B=3, C=1.6, u=0.5$$
\end{center}
\end{figure}

Now $M$ vs $z$ is drawn in figures 3a(i), 3a(ii) and 3a(iii). Since our solution for extended Chaplygin gas model produces only quintessence, so from the figures, we can say that the mass $M$ of the Van der Waal's black hole always increases with decreasing $z$. So we conclude that the mass of the Van der Waal's black hole increases if the extended Chaplygin gas accretes onto the Van der Waal's black hole. Also, if we fix $C$ and vary $\alpha$ then increment of $\alpha$ decreases the value of the mass. However, if we fix $\alpha$ and vary $C$ then increasing $C$ increases the value of the mass. 

For $\alpha=-1$, $M$ can be written as, 
\begin{equation}
M=M_0+\frac{8 \pi u^2}{\sqrt{3}}\left[\left(C+\frac{A}{B-1}\right)^{\frac{1}{2(1-n)}}-\left\{C(1+z)^{3(B-1)(n-1)}+\frac{A}{B-1}\right\}^{\frac{1}{2(1-n)}}\right]
\end{equation}
\begin{figure}[ht]
\begin{center}

~~~~~~~~~~~~~~~~~~~~~~~~~~~~Fig.3b(i)~~~~~~~~~~~~~~~~~~~~~~~~~~~~~~~~~~~~~~~~~\\
\vspace{0.5cm}
\includegraphics[height=2.0in, width=3.0in]{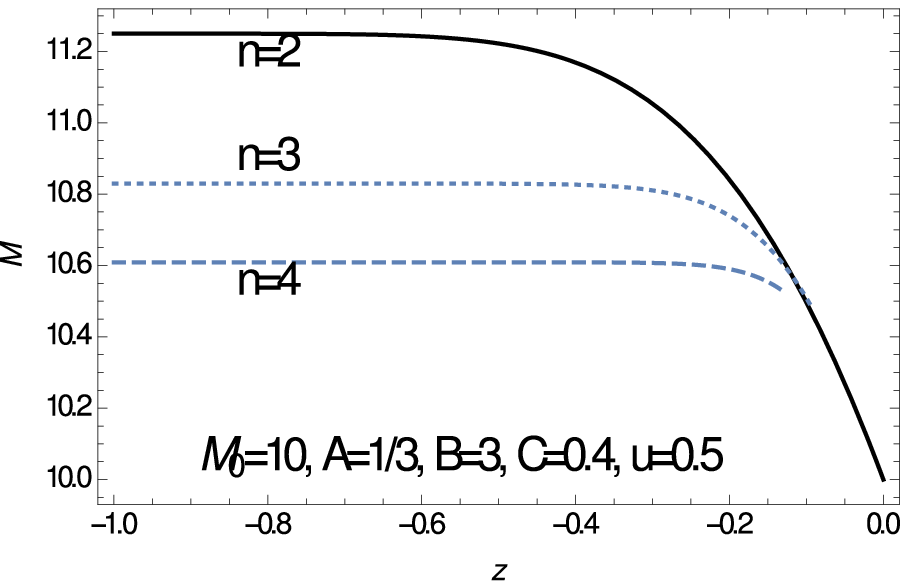}\\
\vspace{0.1cm}
Relation between $M$ and $z$ 
$$M_0=10, A=1/3, B=3, C=0.4, u=0.5$$
\end{center}
\end{figure}

\begin{figure}[ht]
\begin{center}

~~~~~~~~~~~~~~~~~~~~~~~~~~~~Fig.3b(ii)~~~~~~~~~~~~~~~~~~~~~~~~~~~~~~~~~~~~~~~~~\\
\vspace{0.5cm}
\includegraphics[height=2.0in, width=3.0in]{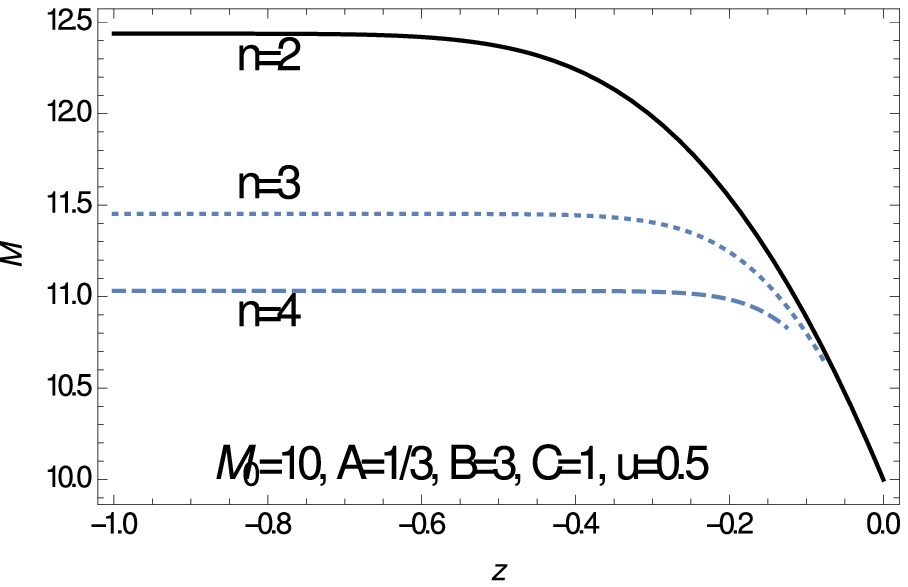}\\
\vspace{0.1cm}
Relation between $M$ and $z$ 
$$M_0=10, A=1/3, B=3, C=1, u=0.5$$
\end{center}
\end{figure}
\begin{figure}[ht]
\begin{center}

~~~~~~~~~~~~~~~~~~~~~~~~~~~~Fig.3b(iii)~~~~~~~~~~~~~~~~~~~~~~~~~~~~~~~~~~~~~~~~~\\
\vspace{0.5cm}
\includegraphics[height=2.0in, width=3.0in]{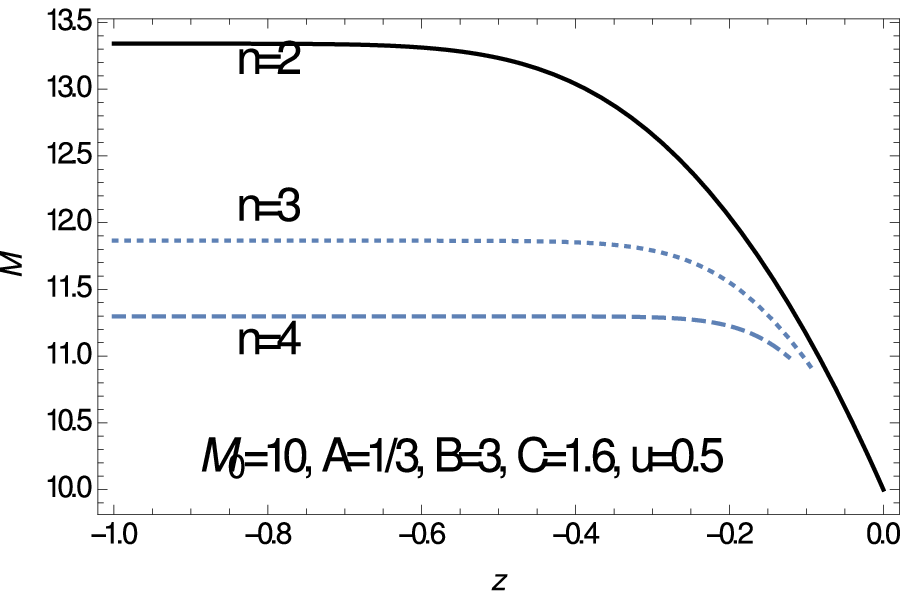}\\
\vspace{0.1cm}
Relation between $M$ and $z$ 
$$M_0=10, A=1/3, B=3, C=1.6, u=0.5$$
\end{center}
\end{figure}

The basic features of the figures 3b(i), 3b(ii) and 3b(iii) do match with the figures 3a(i), 3a(ii) and 3a(iii). So we conclude that the mass of the Van der Waal's black hole increases if the extended Chaplygin gas accretes onto the Van der Waal's black hole. Also, if we fix $C$ and vary $n$, then increment of  $n$ decreases the value of the mass. However, if we fix $n$ and vary $C$ then increment of $C$ increases the value of the mass.

For $n=2$ and $\alpha=\frac{1}{2}$, $M$ can be written as,
\begin{equation}
M=M_0-\frac{8 \pi u^2}{\sqrt{3}}\left[\sqrt{\phi_1(a)}-\sqrt{\phi_1(1)}\right]~~,
\end{equation}
where $\phi_1$ can be evalutated by expressing $\rho$ as a function of scale factor $'a'$ by using the equation (\ref{48}). 

For $n=3$ and $\alpha=\frac{1}{2}$, $M$ can be written as, 
\begin{equation}
M=M_0-\frac{8 \pi u^2}{\sqrt{3}}\left[\sqrt{\phi_2(a)}-\sqrt{\phi_2(1)}\right]~~,
\end{equation}
where $\phi_2$ can be evalutated by expressing $\rho$ as a function of scale factor $'a'$ by using the equation (\ref{48})\\
%%%%%%%%%%%%%%%%%%%%%%%%%%%%%%%%%%%%%%%%%%%%%%%%%%%%%%%%%%%%%%%%%%%%%%%%
\newpage
\section{Discussions}
%%%%%%%%%%%%%%%%%%%%%%%%%%%%%%%%%%%%%%%%%%%%%%%%%%%%%%%%%%%%%%%%%%%%%%%%%%%
In this work, first we have considered the most general static spherically symmetric black hole metric. Then we have studied the accretion onto the Van der Waal's black hole and found some inequalities for physical validation. Next, we have analyzed the thermodynamics of accreting matter around the black hole. We can say that the mass is increasing, i.e., $\dot{M}>0$ in quintessence era, however when we move towards the phantom barrier line, the rate of increment is slowing down, and in phantom era the mass of the black hole is decreasing, i.e., $\dot{M}<0$, which is a point of interest. Finally, we have discussed about dark energy model such as extended Chaplygin gas and the nature of the universe's density. For special case of the modified Chaplygin gas, we have seen that the universe was infinitely dense at its beginning but when the scale factor has turned higher, the universe has started to grow in size. For $\alpha=-1$ in extended Chaplygin gas, the density of accreting fluid is increasing with a steep slope firstly and then the slope is reduced down. In extended Chaplygin gas for $n=2$, $\alpha=\frac{1}{2}$ and $n=3$, $\alpha=\frac{1}{2}$ we obtain an identical nature of the accretion density, i.e., increment in scale factor causes a loss of the density of the accreting fluid. Another reason for decrease of density of the accreting dark energy is increasing mass of the central engine. Since in our solution of modified Chaplygin gas, this model generates only quintessence dark energy and so the Van der Waal's black hole mass increases during the whole evalution of the accelerating universe. We find that the strength of dark energy accretion process on Van der Waal's black hole lies between dark energy accretion on Schwarzschild black hole and adiabatic accretion on Schwarzschild black hole. We speculate that the mutual interactions of microstates of Van der Waal's black hole reduce the outward acting attracting power of it as compared to Schwarzschild black hole when both of them are to attract adiabatic fluid. But when the accreting fluid is Chaplygin gas, due to its negative pressure, Van der Waal's black hole attracts it more than the Schwarzschild one. The accreting power of Van der Waal's black hole lies somewhere between different extremities. It is not so high or not so fainted for different cases like Schwarzschild black hole. 
%%%%%%%%%%%%%%%%%%%%%%%%%%%%%%%%%%%%%%%%%%%%%%%%%%%

\vspace{.25in}
{\bf Acknowledgement:}
This research is supported by the project grant of Goverment of West Bengal, Department of Higher Education, Science and Technology and Biotechnology (File no :- $ST/P/S\&T/16G-19/2017$). SD thanks Goverment of West Bengal, Department of Higher Education, Science and Technology and Biotechnology for non-NET fellowship. RB thanks Inter University Center for Astronomy and Astrophysics(IUCAA), Pune, India for Visiting Associateship. Authors thank Mr. Prasanta Choudhury, M.Sc(Mathematics), Sidhu Kanho Birsa University, West Bengal, India for primary construction of the problem.\\
%%%%%%%%%%%%%%%%%%%%%%%%%%%%%%%%%%%%%%%%%%%%%%%%%

\end{document}